\documentclass[aps,prl,twocolumn,showpacs,superscriptaddress,groupedaddress]{revtex4}  
\usepackage{graphicx}  
\usepackage{dcolumn}   
\usepackage{bm}        
\usepackage{amssymb}   
\usepackage{amsmath}
\usepackage{color}
\usepackage[colorlinks=true, pdfstartview=FitV, linkcolor=blue, citecolor=blue, urlcolor=blue]{hyperref} 
\usepackage{epstopdf}

\usepackage{appendix} 

\newcommand{\ee}[1]{\times 10^{#1}}
\newcommand{\mr}[1]{\mathrm{#1}}

\begin{document}

\preprint{}

\title{Parametric symmetry breaking in a nonlinear resonator}
\author{Anina Leuch}
\thanks{Authors contributed equally.}
\affiliation{Institute for Solid State Physics, ETH Zurich, 8093 Zurich, Switzerland}
\author{Luca Papariello}
\thanks{Authors contributed equally.}
\affiliation{Institute for Theoretical Physics, ETH Zurich, 8093 Zurich, Switzerland}
\author{Oded Zilberberg}
\affiliation{Institute for Theoretical Physics, ETH Zurich, 8093 Zurich, Switzerland}
\author{Christian L. Degen}
\affiliation{Institute for Solid State Physics, ETH Zurich, 8093 Zurich, Switzerland}
\author{R. Chitra}
\affiliation{Institute for Theoretical Physics, ETH Zurich, 8093 Zurich, Switzerland}
\author{Alexander Eichler}
\affiliation{Institute for Solid State Physics, ETH Zurich, 8093 Zurich, Switzerland}
\date{\today}

\begin{abstract}
Much of the physical world around us can be described in terms of harmonic oscillators in thermodynamic equilibrium. At the same time, the far from equilibrium behavior of oscillators is important in many aspects of modern physics. Here, we investigate a resonating system subject to a fundamental interplay between intrinsic nonlinearities and a combination of several driving forces. We have constructed a controllable and robust realization of such a system using a macroscopic doubly clamped string. We experimentally observe a hitherto unseen double hysteresis in both the amplitude and the phase of the resonator's response function and present a theoretical model that is in excellent agreement with the experiment. Our work provides a thorough understanding of the double-hysteretic response through a symmetry breaking of parametric phase states that elucidates the selection criteria governing transitions between stable solutions. Our study motivates applications ranging from ultrasensitive force detection to low-energy computing memory units.
\end{abstract}


\maketitle


Parametric excitation of resonators plays an important role in many areas of science and technology. In its best-known form, parametric excitation describes the modulation of a resonator's natural frequency at twice the natural frequency itself~\cite{Faraday_1831, mathieu1868memoire, rayleigh_1887, Landau_Lifshitz}. In this case, energy is pumped into or out of the resonator depending on the phase of the modulation relative to the oscillation. This ubiquitous feature finds applications in a wide range of fields including signal amplification and noise squeezing~\cite{Penfield_1962, Giordmaine_1965, Caves_1981, Rugar_1991, Castellanos_2007, Karabalin_2011, Koretsky_2012, Devoret_2015, Siddiqi_2015} with contemporary proposals also including topological chiral amplifiers~\cite{peano_2016}, generation of quantum entanglement~\cite{Kwiat_1995, Eichler_2014}, as well as mechanical logic operations with the so-called parametron~\cite{Goto_1959, Mahboob_2008, Mahboob_2011, Mahboob_2014}.

The last decade has seen remarkable progress in the fabrication and control of nanomechanical resonators which serve as an ideal platform for harnessing parametric excitations~\cite{Unterreithmeier_2009, Karabalin_2010, Villanueva_2011}. As the resonators scale down, they attain unprecedented sensitivity towards minute masses, forces and magnetic moments~\cite{RugarSpin, chaste12, moser13}. At the same time, they enter a regime where nonlinearities become a defining characteristic that offers new functionality for parametrical detectors~\cite{PostmaAPL, ZhangNL, Karabalin_2010, Villanueva_2011, Papariello_2016}. Indeed, for sufficiently strong parametric driving, the effective damping of the linear resonator becomes negative and the oscillation amplitude is stabilized by nonlinearities~\cite{Lifshitz_Cross}.

The negative effective damping regime of the parametric resonator is particularly interesting because it features two stable oscillation solutions~\cite{Landau_Lifshitz}. These solutions, which we term `parametric phase states' for the rest of this paper, are a result of the double periodicity of the parametric excitation. They are degenerate in amplitude, but phase shifted by $\pi$, and they are fascinating because they allow for the study of broken time-translation symmetry and activated interstate switching in both classical and quantum systems~\cite{Ryvkine_2006, Chan_2008, Lin_2015}. Recently, it was shown that an external force field can lift the amplitude degeneracy between the parametric phase states~\cite{Papariello_2016}. This degeneracy lifting becomes pronounced in the presence of nonlinear damping and leads to a robust double hysteresis in the frequency-swept response of the resonator, which can be used to measure small near-resonant forces~\cite{Papariello_2016}.

\begin{figure}
\includegraphics[width=\columnwidth]{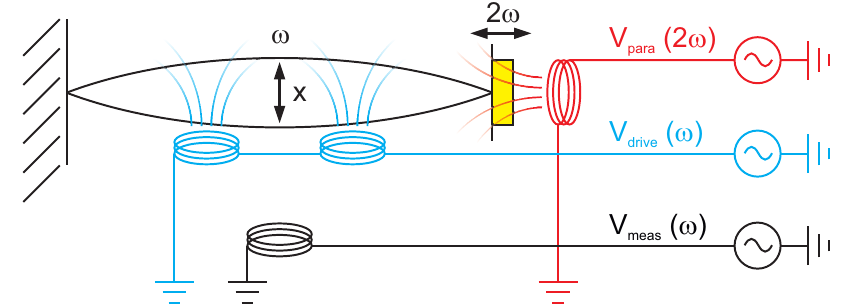}
\caption{\label{fig:figure1} Experimental realization of a parametric resonator based on a doubly clamped steel string ($0.23$\,mm $\times \,0.23$\,mm $\times \,0.36$\,m). Direct driving at frequency $\omega$ and parametric excitation at frequency $2\omega$ rely on AC currents through coils induced by voltages $V_{\rm drive}$ and $V_{\rm para}$, respectively~\cite{supmat}. The string position is read out from the voltage $V_{\rm meas}$ induced in a pickup coil.}
\end{figure}

\begin{figure*}
\includegraphics[width=\textwidth]{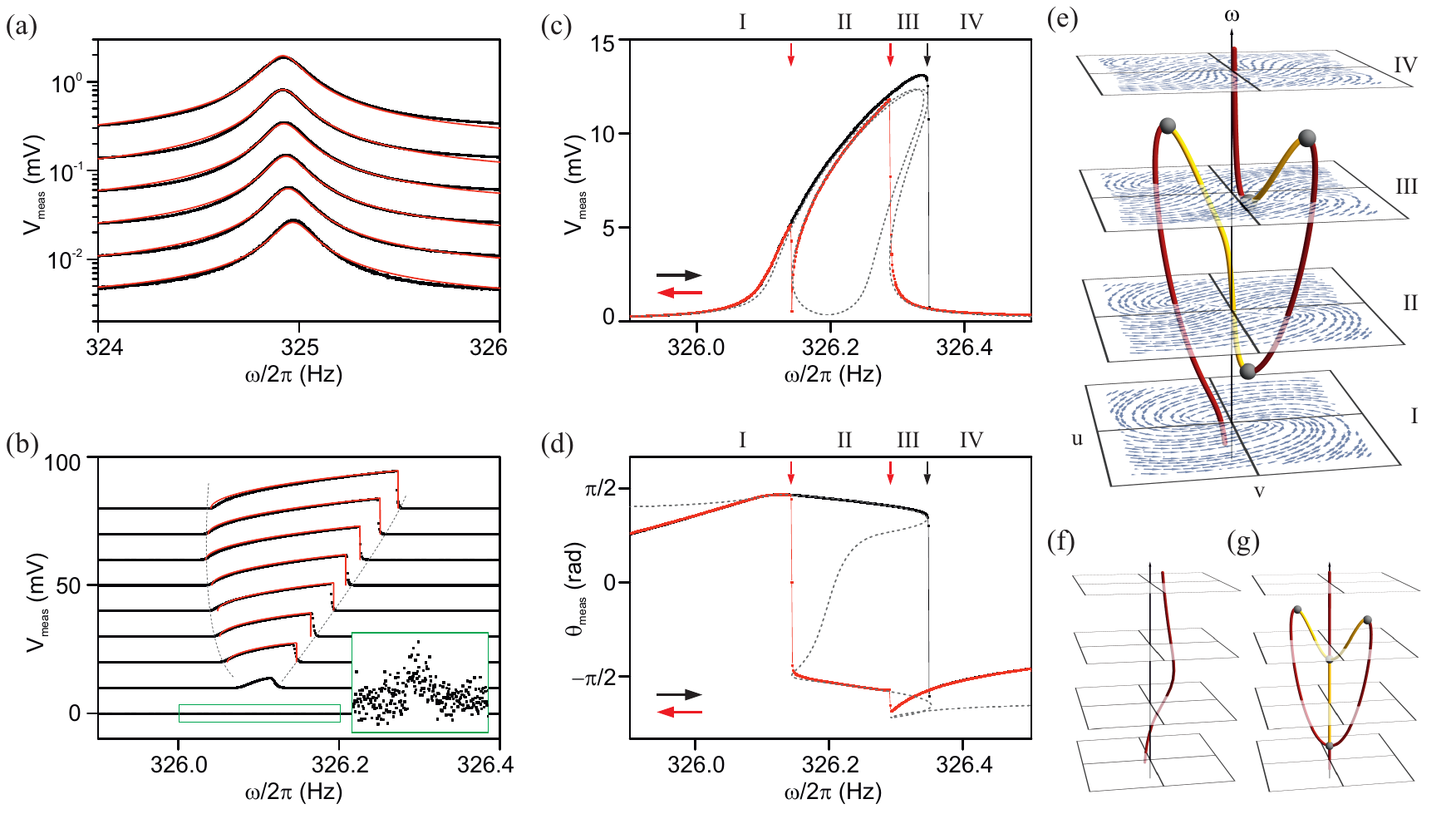}
\caption{\label{fig:figure2}  Device response to the various drives. (a) Linear response with weak external drive amplitudes $V_\mr{drive} = 3.15$\,mV, $7.33$\,mV, $17.1$\,mV, $39.7$\,mV, $92.0$\,mV and $215$\,mV and with $V_\mr{para}=0$. The background increase is due to direct electrical coupling between the drive and detection coils. (b) Response to parametric excitation with $V_\mr{drive} = 0$ and $V_\mr{para} = 0.6-1.0$\,V in steps of $0.05$\,V. Curves are vertically offset by $10$\,mV for better visibility and instability boundaries are traced by gray dashed lines. Inset shows onset of instability for $V_\mr{para} = 0.6$\,V$\equiv V_\mr{th}$. Black dots denote experimental data and all theoretical fits (red solid lines) use $Q = 1800$, $\alpha = 2.45 \ee{10}$\,m$^{-2}$s$^{-2}$ and $\eta = 6.8 \ee{6}$\,m$^{-2}$s$^{-1}$. (c)-(d) First experimental demonstration of double-hysteretic response. (c) mean displacement and (d) oscillation phase as a function of $\omega$ for both upward sweep (black line-dots) and downward sweep (red line-dots). Four domains (I-IV) of the response appear. Here, $V_\mr{drive} = 0.1$\,V, $V_\mr{para}=0.8$\,V and $\phi = -45^{\circ}$. Theory curves are gray dashed lines, cf.~Eqs.~\eqref{eq:slowR} and \eqref{eq:slowP}. (e) For the fitted parameters, representative calculated stability maps of the system in the four domains at $\omega= 0.9997  \omega_0$ in I, $\omega= 1.0 \omega_0$ in II, $\omega= 1.0003 \omega_0$ in III and $\omega= 1.0006 \omega_0$ in IV. Stable solutions (dark red lines) and unstable solutions (bright yellow lines), as well as the bifurcations (grey spheres) as a function of $\omega$ are also shown. In (f) and (g), the corresponding solutions for purely external ($\lambda=0,F\neq 0$) or parametric drives ($\lambda\neq0,F= 0$) are shown. The range of $u=r\cos(\theta)$ and $v=r\sin(\theta)$ axes corresponds to $\pm 0.65$\,mm for (e) and (g), and $\pm 0.065$\,mm for (f). }
\end{figure*}

In this work, we report the first experimental demonstration of the double-hysteretic response and show that it is intimately linked to symmetry breaking between parametric phase states. As a demonstrator, we use a macroscopic mechanical resonator that is similar to state-of-the-art nanomechanical resonators in terms of nonlinear characteristics while offering easy tuning and a signal-to-noise ratio that is rarely attained in nanomechanical devices. We find a complex interplay between driving forces and nonlinearities that leads to multistability in amplitude and phase as a function of driving frequency. In parallel, we present a theoretical model that accurately describes our measurements and lends insight into the governing mechanisms. As an outlook, we describe applications that will profit directly from our study.


Our experimental setup consists of a doubly clamped steel string, see Fig. \ref{fig:figure1}. The string acts as an Euler-Bernoulli beam in the high tension limit~\cite{Lifshitz_Cross}. Parametric excitation is realized by modulation of the position of one clamping point to change the tension inside the string. The motion of the string at angular frequency $\omega$ is transduced into a voltage and read out via a lock-in amplifier~\cite{supmat}. The lowest energy mode of the device satisfies the well-known equation of motion for a nonlinear, parametrically excited resonator~\cite{Lifshitz_Cross}:
\begin{multline} \label{eq:Newton_equation}
\ddot{x} + \omega_0^2 \left[ 1 - \lambda \cos \left(  2 \omega t  \right) \right] x \\+ 
\Gamma \dot{x} + \alpha x^3 + \eta x^2 \dot{x} = \frac{F_0}{M} \cos \left( \omega t  + \phi \right)\,,
\end{multline}
where $x$ is the displacement of the resonator and dots mark differentiations with respect to time $t$. The modulation amplitude $\lambda$ controls the parametric excitation and $\Gamma = \omega_0 / Q$ is the linear damping coefficient with $Q$ the mechanical quality factor. The nonlinearities $\alpha$ and $\eta$ denote the conservative (Duffing-type) and dissipative nonlinearities, respectively. $F_0$ is the amplitude of an applied external force, $M$ is the effective mass of the resonator, which here is equal to half the total mass, and $\phi$ is a phase difference between applied force and parametric excitation~\cite{supmat}.


We use relatively weak external driving to characterize the linear behavior of the device. Figure \ref{fig:figure2}(a) shows the response of the lowest mechanical mode to driving voltages $V_{\rm drive}$ from $3.15$ to $215$\,mV. Optical calibration allows us to translate measured voltage amplitudes, $V_{\rm meas}$, into root mean square displacement, $r$, with a conversion factor of $3.55 \ee {-2}$\,m/$V_{\rm meas}$~\cite{supmat}. We estimate the mass to be $M = 6.5 \ee{-5}$\,kg from the geometry of the string and the density of steel, and fit all response curves with $\omega_0/2\pi \sim 325$\,Hz and $Q = 1800$. The response curves have a purely electrical offset which grows in proportion to $V_\mr{drive}$. For small displacements relative to the string diameter, $r < d= 2.3\ee{-4}$\,m, the peak response is proportional to $V_\mr{drive}$. This allows us to extract a linear relationship between driving voltage and applied force as $F_0 = 4 \ee{-5}$\,N/$V_\mr{drive}$.

To access the nonlinear regime of large displacement amplitudes ($r > d$), we parametrically excite the device. In the absence of an external driving force, we measure large and stable vibrations for values of the parametric excitation voltage $V_\mr{para}$ beyond a threshold of $0.6$\,V, see Fig. \ref{fig:figure2}(b). Using the relationship between linear damping and the parametric instability threshold, $\lambda_\mr{th} = 2/Q$~\cite{Lifshitz_Cross}, we can calibrate the modulation amplitude as $\lambda = \lambda_\mr{th}\cdot V_\mr{para}/ V_\mr{th} \approx V_\mr{para} / 540$\,V. Additionally, we obtain the nonlinear parameters $\alpha = 2.45 \ee{10}$\,m$^{-2}$s$^{-2}$ and $\eta = 1.73 \ee{6}$\,m$^{-2}$s$^{-1}$ by fitting all response curves with the well-known solution of the homogeneous case of Eq.~\eqref{eq:Newton_equation}, with $F_0=0$~\cite{Lifshitz_Cross}. Our fitting relies on the fact that the nonlinear response maps the edges of the so-called ``Arnold's tongue'', i.e., the instability boundaries of the linear parametric resonator, see dashed lines in  Fig. \ref{fig:figure2}(b).

A striking interplay unfolds when parametric excitation and external driving act simultaneously. In Fig. \ref{fig:figure2}(c), the measured displacement amplitude for an upward frequency sweep exhibits a single jump (at the boundary between domains III and IV), akin to the jump expected in standard externally driven Duffing resonators in the absence of parametric excitation. However, for downward frequency sweeps, a double hysteresis appears and the response displays two consecutive jumps (at the III-II and II-I boundaries, respectively). While the jumps (III-IV) and (III-II) describe the typical hysteresis  for externally driven Duffing resonators, the second jump (II-I) is a novel feature that stems from an interplay with parametric excitation and has not been seen before in an experiment. The same hysteretic responses are more prominent in the measured oscillation phase $\theta_\mr{meas}$, see Fig. \ref{fig:figure2}(d).

For a detailed investigation of these features seen in the amplitude and phase measurements,  we solve \eqref{eq:Newton_equation} using the approach discussed in Ref.~\cite{Papariello_2016}. Rewriting Eq.~\eqref{eq:Newton_equation} as two coupled first order differential equations, we use the averaging method to obtain equations of motion for the slow-flow displacement amplitude, $r$, and phase, $\theta$, of the motion

\begin{align}
&\resizebox{.9\hsize}{!}{$\displaystyle\dot{r}  = - \frac{\alpha r \omega \left( 4 \Gamma + \eta r^2 \right) + 2 \alpha k \lambda r \sin 2 \theta + 4 F_0 \alpha \sin \left( \theta - \phi \right)}{8 \omega k \sqrt{\alpha M}} \,, $}
\label{eq:slowR}\\
&\resizebox{.9\hsize}{!}{$\displaystyle\dot{\theta} = \frac{\omega_0}{2 \omega} \left[  \frac{3 \alpha r^2}{4 k} + 1 - \frac{\omega^2}{\omega_0^2} - \frac{\lambda}{2} \cos 2 \theta -\frac{F_0}{k r}{\cos\left( \theta - \phi \right) } \right] \,,$}
\label{eq:slowP}
\end{align}
where $k = M \omega_0^2$. The steady state response is obtained by setting $\dot{r}=\dot{\theta}\equiv0$ and solving the resulting coupled equations. In the presence of both drives, the number of physical solutions varies from one to five depending on $\omega$~\cite{Papariello_2016}. Using the experimentally extracted values for the resonator parameters, we find that the model results, which are shown in Figs.~\ref{fig:figure2}(c) and \ref{fig:figure2}(d), are in excellent agreement with the experiment and allow an unambiguous interpretation of the measured phenomena.  

In Fig. \ref{fig:figure2}(e), we plot the calculated $\omega$-dependent stability diagram of the system. It shows the basins of attraction of the system for domains I-IV along with the evolution of stable attractors and unstable saddle points. One can see clearly how, as a function of $\omega$, the total number of stationary solutions increases (decreases) due to generation (annihilation) of pairs of stable-unstable solutions at bifurcation points (see grey spheres in figure). Correspondingly, in each domain, we have a different number of solutions, i.e., a single solution in I, three in II, five and three in III, and one in IV. Following the evolution of the stable solutions with increasing or decreasing $\omega$ reveals the origin of the second hysteretic jump in both amplitude and phase.

It is instructive to compare the stability diagram in Fig. \ref{fig:figure2}(e) to its two limiting cases, namely the system in the presence of purely external driving or purely parametric excitation. The corresponding stability diagrams are shown in Figs. \ref{fig:figure2}(f) and \ref{fig:figure2}(g), respectively. Intuitively, we can construct the full stability diagram from the purely parametric case by regarding the external drive as a perturbation. As a consequence of this perturbation, the parametric phase states are no longer symmetric and the trivial solutions ($r = 0$) seen in Fig. \ref{fig:figure2}(g) are shifted toward finite amplitudes. Importantly, an opposite phase is imprinted on the stationary solution by the external drive in regions I and IV. As a combined result of these two effects, for opposing directions of frequency sweeps, a different parametric phase state is chosen and the double hysteresis is seen.

\begin{figure}
\includegraphics[width=86mm]{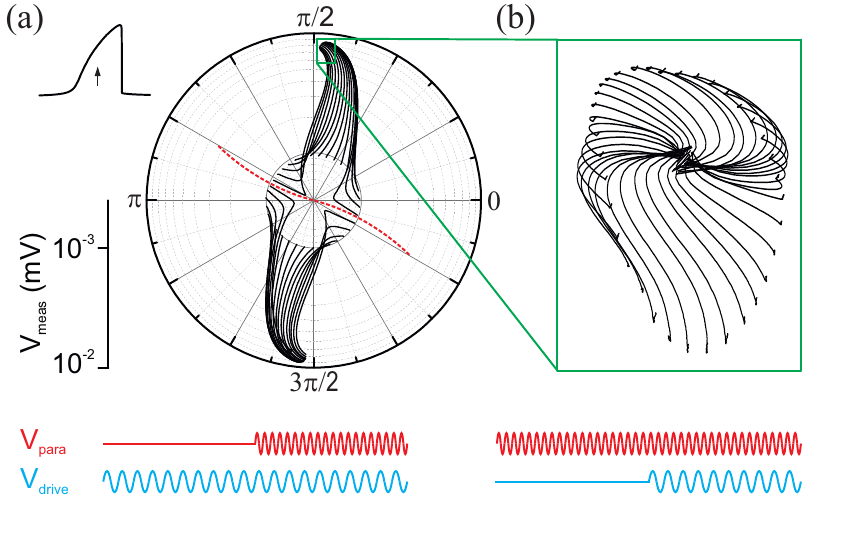}
\caption{\label{fig:figure3} Role of relative phase $\phi$ between drives. Pulse schematics at the bottom illustrate the order of driving/excitation voltages as a function of time for the two experiments. (a) We prepare the resonator at low amplitude $r$ with $V_\mr{drive} = 0.1$\,V at a fixed frequency in domain II (inner ring). When switching on $V_\mr{para} = 0.8$\,V, the resonator rings up to one of the attractors [cf.~Fig.~\ref{fig:figure2}(e)]. Different trajectories use different values of $\phi$. Radial scale is logarithmic. Inset on upper left schematically indicates frequency. (b) Reversing the order of (a), we prepare the resonator with $V_\mr{para} = 0.8$\,V in a pure parametric phase state, then switch on $V_\mr{drive} = 0.1$\,V with varying $\phi$. Here, scale is linear.}
\end{figure}

The central observation of the theoretical analysis above is that the amplitude degeneracy and phase symmetry of parametric phase states are broken by the external driving force. In order to verify this claim, we experimentally probe the stability diagram of the resonator in domain II, see Fig.~\ref{fig:figure3}(a). We prepare the resonator at a fixed frequency and with low amplitude using a small external drive. The phase of the external drive, or equivalently the phase difference $\phi$, determines the starting position of the resonator on the inner circle ($\sim 1$\,mV). Upon activating the parametric excitation, the resonator rings up and settles around one of two attractors, corresponding to one of the perturbed phase states. Clearly, the phase of the starting point determines which attractor is chosen. A red dashed line visualizes the separatrix  stemming from the saddle point associated with the unstable branch, i.e. the boundary between starting points leading to one or the other attractor. Since the imprinted initial phase $\phi$ determines which parametric phase state is chosen, the double hysteresis is visible only for an appropriate range of $\phi$. This is systematically explored in the supplemental material~\cite{supmat}.

To show the  one-to-one correspondence between the pure and perturbed parametric phase states, we prepare the resonator in one of the pure phase states with only parametric excitation, see Fig. \ref{fig:figure3}(b). When the external drive is added, the resonator solution shifts away from - and settles on a ring around - the center of the plot. Again, the final position on the ring depends on $\phi$. Apart from the fact that the perturbed attractors can be mapped onto the original parametric phase states, this experiment also demonstrates that the perturbed phase states are stable for all values of $\phi$, even if the system would preferably select the opposite phase state [as shown in Fig. \ref{fig:figure3}(a)].

\begin{figure}
\includegraphics[width=86mm]{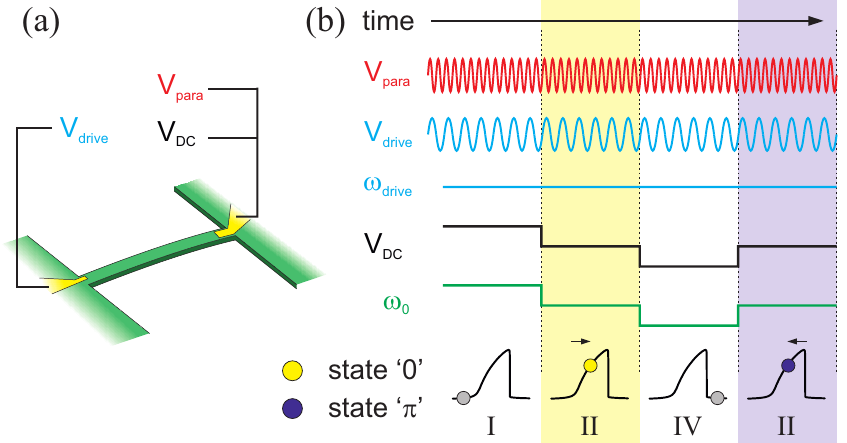}
\caption{\label{fig:figure5} Proposed new parametron control sequence enabled by double hysteresis. (a) Schematic parametric NEMS device with electrodes for drive tones and control voltage. $V_\mr{para}$, $V_\mr{drive}$ and $V_\mr{DC}$ refer to a parametric drive tone applied at $2\omega_\mr{drive}$, an external drive tone at $\omega_\mr{drive}$ and a DC control voltage, respectively. (b) In the proposed control sequence, the driving frequency $\omega_\mr{drive}$ and the relative phase $\phi$ remain fixed, whereas the resonance frequency of the resonator is tuned by $V_\mr{DC}$. Whenever $\omega_\mr{drive}$ coincides with domain II, the resonator switches to one of the parametric phase states `$0$' or `$\pi$' (shaded regions). Which state is chosen depends on the domain in which the resonator was prepared previously.}
\end{figure}

In addition to ultrasensitive force detection as proposed in Ref.~\cite{Papariello_2016}, the double hysteresis opens up new possibilities to control the parametron, a digital storage element that utilizes the parametric phase states to encode bits~\cite{Goto_1959}. Optical and mechanical manifestations of coupled parametrons, where each phase state doublet  plays the role of an artificial spin, are being developed to emulate and solve Ising Hamiltonians~\cite{Mahboob_2016, Inagaki_2016}. In conventional computing, the parametron is a candidate for low-energy consumption memory~\cite{Mahboob_2008}. As we visualize in Fig. \ref{fig:figure5} for the example of a nanomechanical resonator, the double hysteresis allows switching between phase states in a controlled manner without changing amplitude, frequency or phase of the external drive tone (applied at $\omega_\mr{drive}$) and the parametric drive tone (applied at $2\omega_\mr{drive}$). Parametric state switching is performed by changing the resonance frequency of the nanomechanical resonator with a small DC voltage. The parametron resides in one of the two phase states whenever domain II of the resonator overlaps with $\omega_\mr{drive}$. Which phase state is selected depends entirely on whether the resonator was previously prepared in domain I or IV. The parametron can thus be fully controlled by changing a DC voltage. For the device used in Ref.~\cite{Mahboob_2008}, a DC voltage of $\sim 0.2$\,V would be enough to tune the resonator, while the AC drive voltages would be in the $\mu$V range.

The analysis that we present here is valid for any nonlinear resonator subject to a combination of parametric excitation and external driving. Such resonators are actively studied in many modern fields of physics, with examples ranging from levitating nanoparticles~\cite{JanFeedback}, coupled photonic microcavities~\cite{Rodriguez_2016}, nanomechanical resonators and optomechanics to quantum electrodynamics~\cite{DykmanBook}. Future directions include generalization of our results to quantum systems and coupled resonators in the presence of noise.

The authors gratefully acknowledge technical support and know-how from P. M\"arki, C. Keck, U. Grob and T. Ihn. The construction of the setup was done in collaboration with the engineering office (M. Baer) and the mechanical workshop at the Department of Physics at ETH Zurich. This work has been supported by the ERC through Starting Grant 309301, and by the Swiss National Science Foundation.


\begin{thebibliography}{39}
\expandafter\ifx\csname natexlab\endcsname\relax\def\natexlab#1{#1}\fi
\expandafter\ifx\csname bibnamefont\endcsname\relax
  \def\bibnamefont#1{#1}\fi
\expandafter\ifx\csname bibfnamefont\endcsname\relax
  \def\bibfnamefont#1{#1}\fi
\expandafter\ifx\csname citenamefont\endcsname\relax
  \def\citenamefont#1{#1}\fi
\expandafter\ifx\csname url\endcsname\relax
  \def\url#1{\texttt{#1}}\fi
\expandafter\ifx\csname urlprefix\endcsname\relax\def\urlprefix{URL }\fi
\providecommand{\bibinfo}[2]{#2}
\providecommand{\eprint}[2][]{\url{#2}}

\bibitem[{\citenamefont{Faraday}(1831)}]{Faraday_1831}
\bibinfo{author}{\bibfnamefont{M.}~\bibnamefont{Faraday}},
  \bibinfo{journal}{Philosophical transactions of the Royal Society of London}
  \textbf{\bibinfo{volume}{121}}, \bibinfo{pages}{299} (\bibinfo{year}{1831}).

\bibitem[{\citenamefont{Mathieu}(1868)}]{mathieu1868memoire}
\bibinfo{author}{\bibfnamefont{{\'E}.}~\bibnamefont{Mathieu}},
  \bibinfo{journal}{Journal de math{\'e}matiques pures et appliqu{\'e}es}
  \textbf{\bibinfo{volume}{13}}, \bibinfo{pages}{137} (\bibinfo{year}{1868}).

\bibitem[{\citenamefont{Rayleigh}(1887)}]{rayleigh_1887}
\bibinfo{author}{\bibfnamefont{L.}~\bibnamefont{Rayleigh}},
  \bibinfo{journal}{The London, Edinburgh, and Dublin Philosophical Magazine
  and Journal of Science} \textbf{\bibinfo{volume}{24}}, \bibinfo{pages}{145}
  (\bibinfo{year}{1887}).

\bibitem[{\citenamefont{Landau and Lifshitz}(1976)}]{Landau_Lifshitz}
\bibinfo{author}{\bibfnamefont{L.}~\bibnamefont{Landau}} \bibnamefont{and}
  \bibinfo{author}{\bibfnamefont{E.}~\bibnamefont{Lifshitz}},
  \emph{\bibinfo{title}{Mechanics}}, Butterworth-Heinemann
  (\bibinfo{year}{1976}).

\bibitem[{\citenamefont{Penfield and Rafuse}(1962)}]{Penfield_1962}
\bibinfo{author}{\bibfnamefont{P.}~\bibnamefont{Penfield}} \bibnamefont{and}
  \bibinfo{author}{\bibfnamefont{R.~P.} \bibnamefont{Rafuse}},
  \emph{\bibinfo{title}{Varactor Applications}} (\bibinfo{publisher}{MIT Press,
  Cambridge, MA}, \bibinfo{year}{1962}).

\bibitem[{\citenamefont{Giordmaine and Miller}(1965)}]{Giordmaine_1965}
\bibinfo{author}{\bibfnamefont{J.~A.} \bibnamefont{Giordmaine}}
  \bibnamefont{and} \bibinfo{author}{\bibfnamefont{R.~C.}
  \bibnamefont{Miller}}, \bibinfo{journal}{Phys. Rev. Lett.}
  \textbf{\bibinfo{volume}{14}}, \bibinfo{pages}{973} (\bibinfo{year}{1965}).

\bibitem[{\citenamefont{Caves}(1981)}]{Caves_1981}
\bibinfo{author}{\bibfnamefont{C.~M.} \bibnamefont{Caves}},
  \bibinfo{journal}{Phys. Rev. D} \textbf{\bibinfo{volume}{23}},
  \bibinfo{pages}{1693} (\bibinfo{year}{1981}).

\bibitem[{\citenamefont{Rugar and Gr\"utter}(1991)}]{Rugar_1991}
\bibinfo{author}{\bibfnamefont{D.}~\bibnamefont{Rugar}} \bibnamefont{and}
  \bibinfo{author}{\bibfnamefont{P.}~\bibnamefont{Gr\"utter}},
  \bibinfo{journal}{Phys. Rev. Lett.} \textbf{\bibinfo{volume}{67}},
  \bibinfo{pages}{699} (\bibinfo{year}{1991}).

\bibitem[{\citenamefont{Castellanos-Beltran and
  Lehnert}(2007)}]{Castellanos_2007}
\bibinfo{author}{\bibfnamefont{M.~A.} \bibnamefont{Castellanos-Beltran}}
  \bibnamefont{and} \bibinfo{author}{\bibfnamefont{K.~W.}
  \bibnamefont{Lehnert}}, \bibinfo{journal}{Appl. Phys. Lett.}
  \textbf{\bibinfo{volume}{91}}, \bibinfo{pages}{083509}
  (\bibinfo{year}{2007}).

\bibitem[{\citenamefont{Karabalin et~al.}(2011)\citenamefont{Karabalin,
  Lifshitz, Cross, Matheny, Masmanidis, and Roukes}}]{Karabalin_2011}
\bibinfo{author}{\bibfnamefont{R.~B.} \bibnamefont{Karabalin}},
  \bibinfo{author}{\bibfnamefont{R.}~\bibnamefont{Lifshitz}},
  \bibinfo{author}{\bibfnamefont{M.~C.} \bibnamefont{Cross}},
  \bibinfo{author}{\bibfnamefont{M.~H.} \bibnamefont{Matheny}},
  \bibinfo{author}{\bibfnamefont{S.~C.} \bibnamefont{Masmanidis}},
  \bibnamefont{and} \bibinfo{author}{\bibfnamefont{M.~L.}
  \bibnamefont{Roukes}}, \bibinfo{journal}{Phys. Rev. Lett.}
  \textbf{\bibinfo{volume}{106}}, \bibinfo{pages}{094102}
  (\bibinfo{year}{2011}).

\bibitem[{\citenamefont{Qian et~al.}(2012)\citenamefont{Qian, Murphy-Boesch,
  Dodd, and Koretsky}}]{Koretsky_2012}
\bibinfo{author}{\bibfnamefont{C.}~\bibnamefont{Qian}},
  \bibinfo{author}{\bibfnamefont{J.}~\bibnamefont{Murphy-Boesch}},
  \bibinfo{author}{\bibfnamefont{S.}~\bibnamefont{Dodd}}, \bibnamefont{and}
  \bibinfo{author}{\bibfnamefont{A.}~\bibnamefont{Koretsky}},
  \bibinfo{journal}{Magnetic resonance in medicine}
  \textbf{\bibinfo{volume}{68}}, \bibinfo{pages}{989} (\bibinfo{year}{2012}).

\bibitem[{\citenamefont{Sliwa et~al.}(2015)\citenamefont{Sliwa, Hatridge,
  Narla, Shankar, Frunzio, Schoelkopf, and Devoret}}]{Devoret_2015}
\bibinfo{author}{\bibfnamefont{K.}~\bibnamefont{Sliwa}},
  \bibinfo{author}{\bibfnamefont{M.}~\bibnamefont{Hatridge}},
  \bibinfo{author}{\bibfnamefont{A.}~\bibnamefont{Narla}},
  \bibinfo{author}{\bibfnamefont{S.}~\bibnamefont{Shankar}},
  \bibinfo{author}{\bibfnamefont{L.}~\bibnamefont{Frunzio}},
  \bibinfo{author}{\bibfnamefont{R.}~\bibnamefont{Schoelkopf}},
  \bibnamefont{and} \bibinfo{author}{\bibfnamefont{M.}~\bibnamefont{Devoret}},
  \bibinfo{journal}{Physical Review X} \textbf{\bibinfo{volume}{5}},
  \bibinfo{pages}{041020} (\bibinfo{year}{2015}).

\bibitem[{\citenamefont{Macklin et~al.}(2015)\citenamefont{Macklin, Oâ€™Brien,
  Hover, Schwartz, Bolkhovsky, Zhang, Oliver, and Siddiqi}}]{Siddiqi_2015}
\bibinfo{author}{\bibfnamefont{C.}~\bibnamefont{Macklin}},
  \bibinfo{author}{\bibfnamefont{K.}~\bibnamefont{Oâ€™Brien}},
  \bibinfo{author}{\bibfnamefont{D.}~\bibnamefont{Hover}},
  \bibinfo{author}{\bibfnamefont{M.}~\bibnamefont{Schwartz}},
  \bibinfo{author}{\bibfnamefont{V.}~\bibnamefont{Bolkhovsky}},
  \bibinfo{author}{\bibfnamefont{X.}~\bibnamefont{Zhang}},
  \bibinfo{author}{\bibfnamefont{W.}~\bibnamefont{Oliver}}, \bibnamefont{and}
  \bibinfo{author}{\bibfnamefont{I.}~\bibnamefont{Siddiqi}},
  \bibinfo{journal}{Science} \textbf{\bibinfo{volume}{350}},
  \bibinfo{pages}{307} (\bibinfo{year}{2015}).

\bibitem[{\citenamefont{Peano et~al.}(2016)\citenamefont{Peano, Houde,
  Marquardt, and Clerk}}]{peano_2016}
\bibinfo{author}{\bibfnamefont{V.}~\bibnamefont{Peano}},
  \bibinfo{author}{\bibfnamefont{M.}~\bibnamefont{Houde}},
  \bibinfo{author}{\bibfnamefont{F.}~\bibnamefont{Marquardt}},
  \bibnamefont{and} \bibinfo{author}{\bibfnamefont{A.~A.} \bibnamefont{Clerk}},
  \bibinfo{journal}{arXiv preprint arXiv:1604.04179}  (\bibinfo{year}{2016}).

\bibitem[{\citenamefont{Kwiat et~al.}(1995)\citenamefont{Kwiat, Mattle,
  Weinfurter, Zeilinger, Sergienko, and Shih}}]{Kwiat_1995}
\bibinfo{author}{\bibfnamefont{P.~G.} \bibnamefont{Kwiat}},
  \bibinfo{author}{\bibfnamefont{K.}~\bibnamefont{Mattle}},
  \bibinfo{author}{\bibfnamefont{H.}~\bibnamefont{Weinfurter}},
  \bibinfo{author}{\bibfnamefont{A.}~\bibnamefont{Zeilinger}},
  \bibinfo{author}{\bibfnamefont{A.~V.} \bibnamefont{Sergienko}},
  \bibnamefont{and} \bibinfo{author}{\bibfnamefont{Y.}~\bibnamefont{Shih}},
  \bibinfo{journal}{Physical Review Letters} \textbf{\bibinfo{volume}{75}},
  \bibinfo{pages}{4337} (\bibinfo{year}{1995}).

\bibitem[{\citenamefont{Eichler et~al.}(2014)\citenamefont{Eichler, Salathe,
  Mlynek, Schmidt, and Wallraff}}]{Eichler_2014}
\bibinfo{author}{\bibfnamefont{C.}~\bibnamefont{Eichler}},
  \bibinfo{author}{\bibfnamefont{Y.}~\bibnamefont{Salathe}},
  \bibinfo{author}{\bibfnamefont{J.}~\bibnamefont{Mlynek}},
  \bibinfo{author}{\bibfnamefont{S.}~\bibnamefont{Schmidt}}, \bibnamefont{and}
  \bibinfo{author}{\bibfnamefont{A.}~\bibnamefont{Wallraff}},
  \bibinfo{journal}{Phys. Rev. Lett.} \textbf{\bibinfo{volume}{113}},
  \bibinfo{pages}{110502} (\bibinfo{year}{2014}).

\bibitem[{\citenamefont{Goto}(1959)}]{Goto_1959}
\bibinfo{author}{\bibfnamefont{E.}~\bibnamefont{Goto}},
  \bibinfo{journal}{Proceedings of the IRE} \textbf{\bibinfo{volume}{47}},
  \bibinfo{pages}{1304} (\bibinfo{year}{1959}).

\bibitem[{\citenamefont{Mahboob and Yamaguchi}(2008)}]{Mahboob_2008}
\bibinfo{author}{\bibfnamefont{I.}~\bibnamefont{Mahboob}} \bibnamefont{and}
  \bibinfo{author}{\bibfnamefont{H.}~\bibnamefont{Yamaguchi}},
  \bibinfo{journal}{Nature Nanotechnology} \textbf{\bibinfo{volume}{3}},
  \bibinfo{pages}{275} (\bibinfo{year}{2008}).

\bibitem[{\citenamefont{Mahboob et~al.}(2011)\citenamefont{Mahboob, Flurin,
  Nishiguchi, Fujiwara, and Yamaguchi}}]{Mahboob_2011}
\bibinfo{author}{\bibfnamefont{I.}~\bibnamefont{Mahboob}},
  \bibinfo{author}{\bibfnamefont{E.}~\bibnamefont{Flurin}},
  \bibinfo{author}{\bibfnamefont{K.}~\bibnamefont{Nishiguchi}},
  \bibinfo{author}{\bibfnamefont{A.}~\bibnamefont{Fujiwara}}, \bibnamefont{and}
  \bibinfo{author}{\bibfnamefont{H.}~\bibnamefont{Yamaguchi}},
  \bibinfo{journal}{Nat. Commun.} \textbf{\bibinfo{volume}{2}},
  \bibinfo{pages}{198} (\bibinfo{year}{2011}).

\bibitem[{\citenamefont{Mahboob et~al.}(2014)\citenamefont{Mahboob, Mounaix,
  Nishiguchi, Fujiwara, and Yamaguchi}}]{Mahboob_2014}
\bibinfo{author}{\bibfnamefont{I.}~\bibnamefont{Mahboob}},
  \bibinfo{author}{\bibfnamefont{M.}~\bibnamefont{Mounaix}},
  \bibinfo{author}{\bibfnamefont{K.}~\bibnamefont{Nishiguchi}},
  \bibinfo{author}{\bibfnamefont{A.}~\bibnamefont{Fujiwara}}, \bibnamefont{and}
  \bibinfo{author}{\bibfnamefont{H.}~\bibnamefont{Yamaguchi}},
  \bibinfo{journal}{Scientific Reports} \textbf{\bibinfo{volume}{4}},
  \bibinfo{pages}{4448} (\bibinfo{year}{2014}).

\bibitem[{\citenamefont{Unterreithmeier
  et~al.}(2009)\citenamefont{Unterreithmeier, Weig, and
  Kotthaus}}]{Unterreithmeier_2009}
\bibinfo{author}{\bibfnamefont{Q.~P.} \bibnamefont{Unterreithmeier}},
  \bibinfo{author}{\bibfnamefont{E.~M.} \bibnamefont{Weig}}, \bibnamefont{and}
  \bibinfo{author}{\bibfnamefont{J.~P.} \bibnamefont{Kotthaus}},
  \bibinfo{journal}{Nature} \textbf{\bibinfo{volume}{458}},
  \bibinfo{pages}{1001} (\bibinfo{year}{2009}).

\bibitem[{\citenamefont{Karabalin et~al.}(2010)\citenamefont{Karabalin,
  Masmanidis, and Roukes}}]{Karabalin_2010}
\bibinfo{author}{\bibfnamefont{R.}~\bibnamefont{Karabalin}},
  \bibinfo{author}{\bibfnamefont{S.}~\bibnamefont{Masmanidis}},
  \bibnamefont{and} \bibinfo{author}{\bibfnamefont{M.}~\bibnamefont{Roukes}},
  \bibinfo{journal}{Applied Physics Letters} \textbf{\bibinfo{volume}{97}},
  \bibinfo{pages}{183101} (\bibinfo{year}{2010}).

\bibitem[{\citenamefont{Villanueva et~al.}(2011)\citenamefont{Villanueva,
  Karabalin, Matheny, Kenig, Cross, and Roukes}}]{Villanueva_2011}
\bibinfo{author}{\bibfnamefont{L.~G.} \bibnamefont{Villanueva}},
  \bibinfo{author}{\bibfnamefont{R.~B.} \bibnamefont{Karabalin}},
  \bibinfo{author}{\bibfnamefont{M.~H.} \bibnamefont{Matheny}},
  \bibinfo{author}{\bibfnamefont{E.}~\bibnamefont{Kenig}},
  \bibinfo{author}{\bibfnamefont{M.~C.} \bibnamefont{Cross}}, \bibnamefont{and}
  \bibinfo{author}{\bibfnamefont{M.~L.} \bibnamefont{Roukes}},
  \bibinfo{journal}{Nano letters} \textbf{\bibinfo{volume}{11}},
  \bibinfo{pages}{5054} (\bibinfo{year}{2011}).

\bibitem[{\citenamefont{Rugar et~al.}(2004)\citenamefont{Rugar, Budakian,
  Mamin, and Chui}}]{RugarSpin}
\bibinfo{author}{\bibfnamefont{D.}~\bibnamefont{Rugar}},
  \bibinfo{author}{\bibfnamefont{R.}~\bibnamefont{Budakian}},
  \bibinfo{author}{\bibfnamefont{H.~J.} \bibnamefont{Mamin}}, \bibnamefont{and}
  \bibinfo{author}{\bibfnamefont{B.~W.} \bibnamefont{Chui}},
  \bibinfo{journal}{Nature} \textbf{\bibinfo{volume}{430}},
  \bibinfo{pages}{329} (\bibinfo{year}{2004}).

\bibitem[{\citenamefont{Chaste et~al.}(2012)\citenamefont{Chaste, Eichler,
  Moser, Ceballos, Rurali, and Bachtold}}]{chaste12}
\bibinfo{author}{\bibfnamefont{J.}~\bibnamefont{Chaste}},
  \bibinfo{author}{\bibfnamefont{A.}~\bibnamefont{Eichler}},
  \bibinfo{author}{\bibfnamefont{J.}~\bibnamefont{Moser}},
  \bibinfo{author}{\bibfnamefont{G.}~\bibnamefont{Ceballos}},
  \bibinfo{author}{\bibfnamefont{R.}~\bibnamefont{Rurali}}, \bibnamefont{and}
  \bibinfo{author}{\bibfnamefont{A.}~\bibnamefont{Bachtold}},
  \bibinfo{journal}{Nat. Nanotechnol.} \textbf{\bibinfo{volume}{7}},
  \bibinfo{pages}{300} (\bibinfo{year}{2012}).

\bibitem[{\citenamefont{Moser et~al.}(2013)\citenamefont{Moser, Guttinger,
  Eichler, Esplandiu, Liu, Dykman, and Bachtold}}]{moser13}
\bibinfo{author}{\bibfnamefont{J.}~\bibnamefont{Moser}},
  \bibinfo{author}{\bibfnamefont{J.}~\bibnamefont{Guttinger}},
  \bibinfo{author}{\bibfnamefont{A.}~\bibnamefont{Eichler}},
  \bibinfo{author}{\bibfnamefont{M.~J.} \bibnamefont{Esplandiu}},
  \bibinfo{author}{\bibfnamefont{D.~E.} \bibnamefont{Liu}},
  \bibinfo{author}{\bibfnamefont{M.~I.} \bibnamefont{Dykman}},
  \bibnamefont{and} \bibinfo{author}{\bibfnamefont{A.}~\bibnamefont{Bachtold}},
  \bibinfo{journal}{Nature Nanotechnology} \textbf{\bibinfo{volume}{8}},
  \bibinfo{pages}{493} (\bibinfo{year}{2013}).

\bibitem[{\citenamefont{Postma et~al.}(2005)\citenamefont{Postma, Kozinsky,
  Husain, and Roukes}}]{PostmaAPL}
\bibinfo{author}{\bibfnamefont{H.~W.~C.} \bibnamefont{Postma}},
  \bibinfo{author}{\bibfnamefont{I.}~\bibnamefont{Kozinsky}},
  \bibinfo{author}{\bibfnamefont{A.}~\bibnamefont{Husain}}, \bibnamefont{and}
  \bibinfo{author}{\bibfnamefont{M.~L.} \bibnamefont{Roukes}},
  \bibinfo{journal}{Applied Physics Letters} \textbf{\bibinfo{volume}{86}},
  \bibinfo{eid}{223105} (\bibinfo{year}{2005}).

\bibitem[{\citenamefont{Zhang et~al.}(2002)\citenamefont{Zhang, Baskaran, and
  Turner}}]{ZhangNL}
\bibinfo{author}{\bibfnamefont{W.}~\bibnamefont{Zhang}},
  \bibinfo{author}{\bibfnamefont{R.}~\bibnamefont{Baskaran}}, \bibnamefont{and}
  \bibinfo{author}{\bibfnamefont{K.~L.} \bibnamefont{Turner}},
  \bibinfo{journal}{Sensors and Actuators A: Physical}
  \textbf{\bibinfo{volume}{102}}, \bibinfo{pages}{139} (\bibinfo{year}{2002}).

\bibitem[{\citenamefont{Papariello et~al.}(2016)\citenamefont{Papariello,
  Zilberberg, Eichler, and Chitra}}]{Papariello_2016}
\bibinfo{author}{\bibfnamefont{L.}~\bibnamefont{Papariello}},
  \bibinfo{author}{\bibfnamefont{O.}~\bibnamefont{Zilberberg}},
  \bibinfo{author}{\bibfnamefont{A.}~\bibnamefont{Eichler}}, \bibnamefont{and}
  \bibinfo{author}{\bibfnamefont{R.}~\bibnamefont{Chitra}},
  \bibinfo{journal}{Phys. Rev. E} \textbf{\bibinfo{volume}{94}},
  \bibinfo{pages}{022201} (\bibinfo{year}{2016}).

\bibitem[{\citenamefont{Lifshitz}(2009)}]{Lifshitz_Cross}
\bibinfo{author}{\bibfnamefont{M.~C.} \bibnamefont{Lifshitz},
  \bibfnamefont{R.~Cross}}, \emph{\bibinfo{title}{Nonlinear Dynamics of
  Nanomechanical and Micromechanical Resonators}}
  (\bibinfo{publisher}{Wiley-VCH}, \bibinfo{year}{2009}), pp.
  \bibinfo{pages}{1--52}.

\bibitem[{\citenamefont{Ryvkine and Dykman}(2006)}]{Ryvkine_2006}
\bibinfo{author}{\bibfnamefont{D.}~\bibnamefont{Ryvkine}} \bibnamefont{and}
  \bibinfo{author}{\bibfnamefont{M.~I.} \bibnamefont{Dykman}},
  \bibinfo{journal}{Physical Review E} \textbf{\bibinfo{volume}{74}},
  \bibinfo{pages}{061118} (\bibinfo{year}{2006}).

\bibitem[{\citenamefont{Chan et~al.}(2008)\citenamefont{Chan, Dykman, and
  Stambaugh}}]{Chan_2008}
\bibinfo{author}{\bibfnamefont{H.}~\bibnamefont{Chan}},
  \bibinfo{author}{\bibfnamefont{M.}~\bibnamefont{Dykman}}, \bibnamefont{and}
  \bibinfo{author}{\bibfnamefont{C.}~\bibnamefont{Stambaugh}},
  \bibinfo{journal}{Physical review letters} \textbf{\bibinfo{volume}{100}},
  \bibinfo{pages}{130602} (\bibinfo{year}{2008}).

\bibitem[{\citenamefont{Lin et~al.}(2015)\citenamefont{Lin, Nakamura, and
  Dykman}}]{Lin_2015}
\bibinfo{author}{\bibfnamefont{Z.}~\bibnamefont{Lin}},
  \bibinfo{author}{\bibfnamefont{Y.}~\bibnamefont{Nakamura}}, \bibnamefont{and}
  \bibinfo{author}{\bibfnamefont{M.}~\bibnamefont{Dykman}},
  \bibinfo{journal}{Physical Review E} \textbf{\bibinfo{volume}{92}},
  \bibinfo{pages}{022105} (\bibinfo{year}{2015}).

\bibitem[{sup()}]{supmat}
\bibinfo{note}{For additional experimental details, see Supplemental Material.}

\bibitem[{\citenamefont{Mahboob et~al.}(2016)\citenamefont{Mahboob, Okamoto,
  and Yamaguchi}}]{Mahboob_2016}
\bibinfo{author}{\bibfnamefont{I.}~\bibnamefont{Mahboob}},
  \bibinfo{author}{\bibfnamefont{H.}~\bibnamefont{Okamoto}}, \bibnamefont{and}
  \bibinfo{author}{\bibfnamefont{H.}~\bibnamefont{Yamaguchi}},
  \bibinfo{journal}{Science Advances} \textbf{\bibinfo{volume}{2}}
  (\bibinfo{year}{2016}).

\bibitem[{\citenamefont{Inagaki et~al.}(2016)\citenamefont{Inagaki, Inaba,
  Hamerly, Yamamoto, and Takesue}}]{Inagaki_2016}
\bibinfo{author}{\bibfnamefont{T.}~\bibnamefont{Inagaki}},
  \bibinfo{author}{\bibfnamefont{K.}~\bibnamefont{Inaba}},
  \bibinfo{author}{\bibfnamefont{K.}~\bibnamefont{Hamerly},
  \bibfnamefont{R.~andInoue}},
  \bibinfo{author}{\bibfnamefont{Y.}~\bibnamefont{Yamamoto}}, \bibnamefont{and}
  \bibinfo{author}{\bibfnamefont{H.}~\bibnamefont{Takesue}},
  \bibinfo{journal}{Nature Photonics} \textbf{\bibinfo{volume}{10}},
  \bibinfo{pages}{415} (\bibinfo{year}{2016}).

\bibitem[{\citenamefont{Gieseler et~al.}(2012)\citenamefont{Gieseler, Deutsch,
  Quidant, and Novotny}}]{JanFeedback}
\bibinfo{author}{\bibfnamefont{J.}~\bibnamefont{Gieseler}},
  \bibinfo{author}{\bibfnamefont{B.}~\bibnamefont{Deutsch}},
  \bibinfo{author}{\bibfnamefont{R.}~\bibnamefont{Quidant}}, \bibnamefont{and}
  \bibinfo{author}{\bibfnamefont{L.}~\bibnamefont{Novotny}},
  \bibinfo{journal}{Phys. Rev. Lett.} \textbf{\bibinfo{volume}{109}},
  \bibinfo{pages}{103603} (\bibinfo{year}{2012}).

\bibitem[{\citenamefont{Rodriguez et~al.}(2011)\citenamefont{Rodriguez, Amo,
  Sagnes, Le~Gratinet, Galopin, Lema\^{i}tre, and Bloch}}]{Rodriguez_2016}
\bibinfo{author}{\bibfnamefont{S.}~\bibnamefont{Rodriguez}},
  \bibinfo{author}{\bibfnamefont{A.}~\bibnamefont{Amo}},
  \bibinfo{author}{\bibfnamefont{I.}~\bibnamefont{Sagnes}},
  \bibinfo{author}{\bibfnamefont{L.}~\bibnamefont{Le~Gratinet}},
  \bibinfo{author}{\bibfnamefont{E.}~\bibnamefont{Galopin}},
  \bibinfo{author}{\bibfnamefont{A.}~\bibnamefont{Lema\^{i}tre}},
  \bibnamefont{and} \bibinfo{author}{\bibfnamefont{J.}~\bibnamefont{Bloch}},
  \bibinfo{journal}{Nano letters} \textbf{\bibinfo{volume}{11}},
  \bibinfo{pages}{5054} (\bibinfo{year}{2011}).

\bibitem[{\citenamefont{Dykman}(2012)}]{DykmanBook}
\bibinfo{author}{\bibfnamefont{M.}~\bibnamefont{Dykman}},
  \emph{\bibinfo{title}{Fluctuating Nonlinear Oscillators}}
  (\bibinfo{publisher}{Oxford University Press}, \bibinfo{year}{2012}).

\end{thebibliography}

\end{document}